\documentclass[10pt,conference]{IEEEtran}
\IEEEoverridecommandlockouts

\usepackage[utf8]{inputenc} 
\usepackage[T1]{fontenc}    
\usepackage[colorlinks=true, allcolors=black]{hyperref}
\usepackage{url}            
\usepackage{booktabs}       
\usepackage{amsfonts}       
\usepackage{nicefrac}       
\usepackage{microtype}      
\usepackage{xcolor}         

\usepackage{cite}
\usepackage{graphicx}
\usepackage{float}
\usepackage{xspace}
\usepackage{tabularx}
\usepackage{longtable} 
\usepackage{colortbl}
\usepackage{siunitx}
\usepackage{multirow}
\usepackage{makecell}
\usepackage{dcolumn}
\usepackage{multirow}
\usepackage{pifont}
\usepackage{placeins}
\usepackage{tikz}
\usepackage{amsmath}
\usepackage{enumitem}
\usepackage{subfigure}
\usepackage[most]{tcolorbox}
\usepackage[colorinlistoftodos,prependcaption]{todonotes}

\makeatletter
\newcommand{\DeclareLatinAbbrev}[2]{%
  \DeclareRobustCommand{#1}{%
    \@ifnextchar{.}{\textit{#2}}{%
      \@ifnextchar{,}{\textit{#2.}}{%
        \@ifnextchar{!}{\textit{#2.}}{%
          \@ifnextchar{?}{\textit{#2.}}{%
            \@ifnextchar{)}{\textit{#2.}}{%
              {\textit{#2.,\ }}}}}}}}%
}
\makeatother
\DeclareLatinAbbrev{\eg}{e.g}
\DeclareLatinAbbrev{\Eg}{E.g}
\DeclareLatinAbbrev{\ie}{i.e}
\DeclareLatinAbbrev{\Ie}{I.e}
\DeclareLatinAbbrev{\etc}{etc}
\DeclareLatinAbbrev{\etal}{et~al}

\def\first {$(i)$\xspace}
\def\second{$(ii)$\xspace}
\def\third {$(iii)$\xspace}

\makeatletter
    \newcommand{\linebreakand}{%
      \end{@IEEEauthorhalign}
      \hfill\mbox{}\par
      \mbox{}\hfill\begin{@IEEEauthorhalign}
    }
\makeatother

\newcommand{\smartparagraph}[1]{\vspace{.05in}\noindent\textbf{#1}}

\def\BibTeX{{\rm B\kern-.05em{\sc i\kern-.025em b}\kern-.08em
    T\kern-.1667em\lower.7ex\hbox{E}\kern-.125emX}}
\begin{document}

\bstctlcite{IEEEexample:BSTcontrol}

\title{Automating the Detection of \\Code Vulnerabilities by Analyzing GitHub Issues
}
\author{\IEEEauthorblockN{Daniele Cipollone$^{\dagger}$\thanks{
$^{\dagger}$Work performed while at RISE AB.}}
\IEEEauthorblockA{\textit{Delft University of Technology}}
\and
\IEEEauthorblockN{Changjie Wang}
\IEEEauthorblockA{\textit{KTH Royal Institute of Technology} \\
}
\and
\IEEEauthorblockN{Mariano Scazzariello}
\IEEEauthorblockA{\textit{RISE AB} \\
}
\and
\IEEEauthorblockN{Simone Ferlin}
\IEEEauthorblockA{\textit{Red Hat} \\
}
\linebreakand
\IEEEauthorblockN{Maliheh Izadi}
\IEEEauthorblockA{\textit{Delft University of Technology} \\
}
\and
\IEEEauthorblockN{Dejan Kosti\'c}
\IEEEauthorblockA{\textit{KTH Royal Institute of Technology} \\
\textit{RISE AB}}
\and
\IEEEauthorblockN{Marco Chiesa}
\IEEEauthorblockA{\textit{KTH Royal Institute of Technology} \\
}
}


\maketitle

\begin{abstract}
In today's digital landscape, the importance of timely and accurate vulnerability detection has significantly increased. This paper presents a novel approach that leverages transformer-based models and machine learning techniques to automate the identification of software vulnerabilities by analyzing GitHub issues. We introduce a new dataset specifically designed for classifying GitHub issues relevant to vulnerability detection. We then examine various classification techniques to determine their effectiveness. The results demonstrate the potential of this approach for real-world application in early vulnerability detection, which could substantially reduce the window of exploitation for software vulnerabilities. This research makes a key contribution to the field by providing a scalable and computationally efficient framework for automated detection, enabling the prevention of compromised software usage before official notifications. This work has the potential to enhance the security of open-source software ecosystems.

\end{abstract}

\section{Introduction}

With increasing pressure on faster software development and continuous updates, the risk of overseeing bugs or vulnerabilities increases. As such, software refinement and repair becomes increasingly critical. Developers work to address vulnerabilities as they are discovered using various methods, ranging from low-level tests on executed code to simulations of potential attack scenarios to prevent system disruption. Despite enormous efforts, it remains challenging to entirely eliminate vulnerabilities, especially in large code bases. Among these, \textit{zero-day vulnerabilities} pose a particularly severe threat. They are flaws in applications (or even in operating systems) that still have not been discovered by the software manufacturer and, as a result, lack corresponding patches. Google's Threat Analysis Group reported a significant increase in the exploitation of zero-day vulnerability in March 2024, emphasizing the need for more proactive threat detection~\cite{noauthor_review_2024}.

The Common Vulnerabilities and Exposures (CVE) system, managed by the MITRE Corporation~\cite{cve-main}, plays a crucial role in proactive and public vulnerability detection and classification. It offers a standardized framework for identifying, cataloging, and disclosing vulnerabilities. To ensure that organizations and tools consistently refer to the same vulnerability, the CVE system assigns a unique identifier to each reported vulnerability through a well-defined process~\cite{lin_coordination_2023}. Thus, each CVE entry includes a unique identifier, a concise description of the vulnerability, and associated metadata such as affected products, severity levels, and potential impacts. On the other hand, the software industry commonly adopts the term \textit{embargo} for internal CVEs, to restrict or prohibit communication about vulnerabilities~\cite{redhat-embargo}. The \textit{embargo period} describes the time that the flaw is known privately (prior to a deadline, \eg \textit{go-live}). Since their impact can be very unpredictable to several parties, embargoed flaws are very time-critical and require to be handled responsibly, thereby reducing the risk of disclosure in GitHub issues or PRs, public Bugzilla or Jira, public mailing-lists, or unnecessary broad communication. After the fix can be made public, the embargo is thus lifted. 

\smartparagraph{Hints about vulnerabilities may be involuntarily disclosed before the official CVE disclosure.} 
Current methodologies for vulnerability detection rely on a combination of community reporting and proactive research conducted by private, specialized security research services. The process significantly involves human expertise to identify vulnerabilities and relies on responsible disclosure, \textit{i.e.}, a vulnerability is made public only after the responsible parties have been given time to apply a patch. Nevertheless, information about vulnerabilities may still be shared through informal channels (\eg mailing lists, GitHub issues, and blogs)~\cite{sauerwein_shadow_2018}, even before an official disclosure~\cite{syed_what_2018}. In these contexts, \textit{issues may initially be perceived as software bugs or malfunctions rather than recognized as vulnerabilities}. For instance, vulnerability CVE-2016-5696
describes an issue in the Linux kernel that allows the hijacking of TCP, and was discussed on social media up to one month before its release in the CVE database~\cite{sauerwein_tweet_2018}. 

\smartparagraph{The CVE creation process is lengthy.} The process from vulnerability discovery to the official assignment of a CVE can be time-consuming, as it involves rigorous validation to confirm that the vulnerability meets the criteria for CVE classification. For example, in the case of CVE-2023-48305~\cite{noauthor_cve_nodate}, nearly six months passed between the initial discussions of the bug in the GitHub issues to the official creation of the CVE entry. This delay significantly affects end-users, as they may remain unaware of a critical vulnerability and continue using the compromised software without any official notification.


\newpage
\smartparagraph{Can Transformer-based models detect early indicators of a potential CVE vulnerability disclosure?} Large Language Models (LLMs) are Transformer-based AI models trained on vast amounts of text, enabling them to understand and interpret human language, excelling in recognizing text patterns and extracting critical information~\cite{tornberg_how_2023, petukhova_text_2024}. Unlike other ML models, LLMs can effectively process extensive human-written documents thanks to their large context windows. Also, LLMs have the capability to analyze and generate code across various programming languages, as demonstrated by their performance across multiple benchmarks~\cite{evalplus}. This makes LLMs highly suitable for vulnerability detection, as they can automatically analyze content from several, different sources like social media, blogs, mailing lists, and GitHub issues to identify and report potential vulnerabilities to developers.

\smartparagraph{Scope of the paper.} In this paper, we explore the potential of Transformer-based models to identify software vulnerabilities through the analysis of communication channels, specifically focusing on GitHub issues.\footnote{We plan to extend the work to mailing lists and blogs in the future.} This work aims to determine whether it is feasible to identify early indicators of software vulnerabilities with existing models. Our findings not only demonstrate the viability of these approaches, but also highlight their potential to greatly improve current cybersecurity practices by facilitating earlier vulnerability detection, thereby reducing the risk of exploitation.

\smartparagraph{Contributions.} We make three contributions:
\begin{itemize}[leftmargin=*,noitemsep]
    \item We are the first to build a dataset that classifies GitHub issues into those connected with CVEs and those that are not, associating the latter with their corresponding CVE records. 
    \item To the best of our knowledge, using our dataset, we are the first ones to investigate whether Transformer-based models (\ie LLMs and embedding models) can effectively detect undisclosed vulnerabilities from GitHub issues.
    \item Our findings validate the effectiveness of Transformer-based models in this context, with accuracy $\sim$2$\times$ higher compared to statistical approaches. These results highlight their potential for early-stage vulnerability detection.
\end{itemize}

\section{Related Work}

\smartparagraph{Early vulnerability detection.} Detecting vulnerabilities before their public disclosure remains a largely unexplored research area. The study ``The Tweet Advantage''~\cite{sauerwein_tweet_2018} examines $\sim$709\,K tweets that include CVE identifiers posted between May 23, 2016, and March 27, 2018, mapping them to a vulnerability lifecycle model. Their results indicate that nearly one-quarter of vulnerabilities were mentioned on Twitter prior to their official disclosure by vendors or authoritative sources. While the study underscores Twitter's potential as an early warning system for newly discovered vulnerabilities, it relies on simple regexes to detect CVE IDs within the text that indicate vulnerability descriptions.
%
Previous work explored the use of deep learning techniques to assess vulnerability severity based on textual descriptions. For example, Han et al.~\cite{related_work_2} integrate a word embedding layer with a Convolutional Neural Network (CNN) layer to automatically identify critical sentences in vulnerability descriptions and predict their severity. However, this approach relies on humans to first identify and extract relevant vulnerability descriptions, which presents additional challenges for developers who must manually filter input data before applying the model. Instead, we aim to build a system that autonomously selects and analyses the relevant input data. 

\smartparagraph{LLMs for vulnerability detection.} Although the application of LLMs for vulnerability detection is not novel, most existing research focuses on identifying potential bug risks within code snippets~\cite{LLM4v_1, LLM4v_2, LLM4v_3, LLM4v_4}. These methods are designed to detect vulnerabilities within well-defined contexts, such as through static code analysis. In contrast, our paper introduces a new approach by leveraging LLMs to detect vulnerabilities that have not yet been officially disclosed, using textual sources such as social media and GitHub issues.

\section{A Dataset for Vulnerability Detection}\label{sec:dataset}

Before applying different Transformer-based models, it is necessary to identify a suitable dataset for the task. Although previous research has explored the use of LLMs for vulnerability detection, these studies have primarily focused on static code analysis, hence, lacking any textual-based dataset derived from mailing lists or similar sources.

Therefore, we are the first to create a dataset that directly links vulnerability-related textual discussions from GitHub issues in open-source projects with their corresponding entries in the CVE database.

\vspace{.05in}
In the following, we provide a detailed explanation of the methodology used for the dataset creation. We first outline the data source chosen for extraction, followed by a description of the data collection pipeline.


\smartparagraph{Data source.} We aim to build a dataset that determines whether a specific textual description of an unexpected behavior is associated with a potential vulnerability. To achieve this, we require a vast amount of textual data paired with a ground truth indicating whether the text pertains to a disclosed vulnerability or not.
%
We rely on the National Vulnerability Database (NVD), the U.S. government's repository of standards based vulnerability management data~\cite{nvd}. NVD offers a standardized platform for systematically identifying and tracking CVE-based vulnerabilities. It collects comprehensive information on vulnerabilities, including CVE identifiers, affected software versions, severity ratings, and potential impact assessments. Most importantly, each NVD entry provides links to related external resources, \eg issues tracking (on GitHub or similar websites), release notes, and vendor advisories.

\begin{table}[t]
\caption{Top 10 Web Domains in CVE External References}
\label{tab:top_domains}
\centering
\small 
\begin{tabular}{c c}
\textbf{Domain} & \textbf{Count} \\
\hline
    github.com & 42,165 \\
    git.kernel.org & 10,461 \\
    vuldb.com & 8,332 \\
    lists.fedoraproject.org & 4,790 \\
    patchstack.com & 4,256 \\
    lists.apache.org & 3,805 \\
    www.zerodayinitiative.com & 3,620 \\
    packetstormsecurity.com & 3,523 \\
    portal.msrc.microsoft.com & 3,340 \\
    plugins.trac.wordpress.org & 3,258
\end{tabular}
\end{table}

\smartparagraph{Data collection.} 
%
We initially extracted all the vulnerabilities published in the NVD from January 1st, 2019, to June 2nd, 2024. The extracted data include key information such as CVE identifiers, vulnerability descriptions, and severity metrics.
We excluded vulnerabilities that were under examination or had been rejected from the dataset, resulting in a total of 113,735 entries. For each valid vulnerability, we extracted the list of external references and built a comprehensive dataset that links each reference to its corresponding CVE-ID. Table~\ref{tab:top_domains} shows the distribution of the top 10 web domains referenced. We can infer that GitHub is one of the most essential resources for defining and understanding vulnerabilities. Hence, we selected GitHub issues as our main source for textual data.

Among the extracted references, we identified a total of 6,626 distinct GitHub repositories within the considered time frame. Like any code repository, the majority of the issues were unrelated to any vulnerability. To create a high-quality dataset with a significant portion of vulnerabilities, we refined our selection by ranking the repositories based on the number of associated vulnerabilities over the past year and selecting only the top 31, thus ensuring that only the most relevant repositories for vulnerability tracking were included. For each selected repository, we mark the GitHub issues cited in the CVE external references, as these serve as our ground truth. Also, other issues from the same repository are proportionally included to enrich the dataset with samples not related to vulnerabilities. We excluded issues exceeding 8,191 tokens. This limit aligns with the context window size of most of the open-source LLMs (\ie less than 8,192 tokens).

\smartparagraph{The final dataset.} Following data collection and processing, we have created the first dataset comprising a total of 4,379 GitHub issues. Of these, 844 are linked to disclosed CVE vulnerabilities, while 3,535 represent issues unrelated to any vulnerabilities. These issues are taken from 31 different repositories. For each issue linked to a vulnerability, we include CVE ID, disclosure date, vulnerability description, a list of references, details on the affected software, and vulnerability impact metrics updated until August 11th, 2024.



\section{Transformer-based Vulnerability Detection}\label{sec:pipelines}

In this section, we explore the potential of using Transformer-based models for early vulnerability detection through textual data, \ie GitHub issues. Leveraging the dataset built in Sec.~\ref{sec:dataset}, we design three different approaches to detect vulnerabilities: \first an \textit{embedding-based classifier} employing XGBoost, \second an \textit{LLM-based detection} utilizing GPT-3.5 (both base and fine-tuned versions), and \third a \textit{combined approach} that integrates both LLMs and XGBoost.

\begin{figure}[b]
    \centering
    \includegraphics[width=\linewidth]{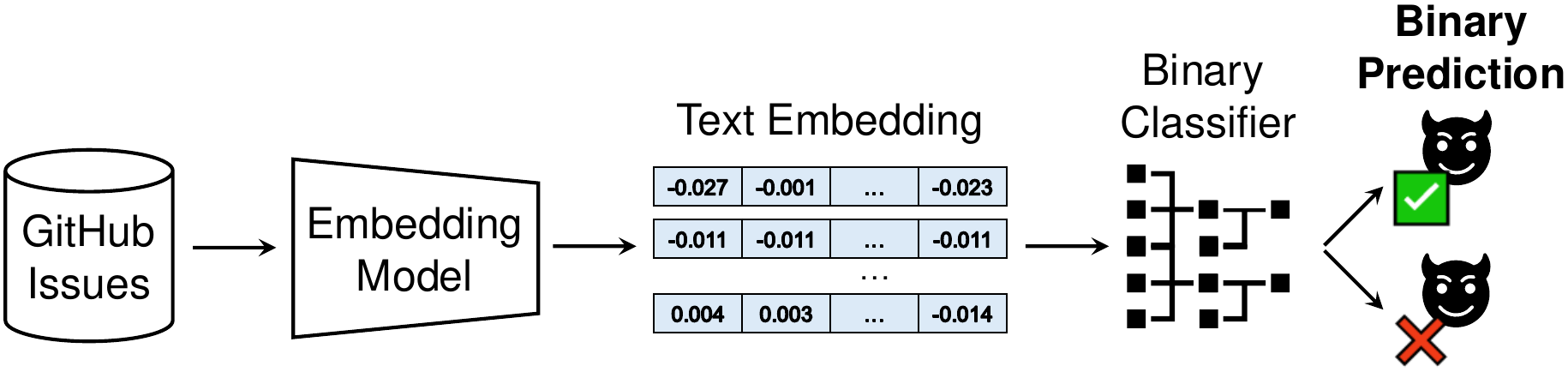}
  \caption{Embedding-based XGBoost classifier.}
  \label{fig:embedding_classfier}
\end{figure}

\subsection{Embedding-based Vulnerability Classifier}\label{ss:embedding}
In the first approach, we do not directly rely on LLMs but instead use embedding models to train a task-specific classifier. Embedding models convert text into dense, high-dimensional vectors that capture semantic relationships and contextual nuances~\cite{bert}. Modern embedding models are often derived from pre-trained LLMs~\cite{nvembed,oaiembed}, as they can leverage the deep contextual comprehension and rich semantic representations acquired during their training. This leads to more precise and meaningful embeddings for downstream applications. 
%

In our approach, depicted in Fig.~\ref{fig:embedding_classfier}, we begin by representing each GitHub issue (using the format shown in Fig.~\ref{fig:prompt}) as a vector in a continuous space, allowing us to identify patterns that may not be visible in the raw text. These embeddings are then fed as input features for a classifier, which we trained to produce a binary prediction on whether a given GitHub issue is associated with a vulnerability or not.

\smartparagraph{Model selection.} For embedding generation, we selected two models: the proprietary \texttt{text-embedding-3-large} from OpenAI~\cite{noauthor_new_nodate}, and the open-source \texttt{NV-Embed-v2} from Nvidia~\cite{nvembed}, which, at the time of writing, is the leading model on the Massive Text Embedding Benchmark (MTEB) leaderboard~\cite{muennighoff2022mteb}. \texttt{NV-Embed-v2} is derived from \texttt{Mistral-7B-v0.1}, while there is no public information available on the architecture and training of \texttt{text-embedding-3-large}.
For the classification task, we chose the XGBoost model~\cite{chen_xgboost_2016}, a powerful gradient-boosting algorithm well-known for its strong performance in classification tasks. We selected XGBoost due to its capability to handle large datasets, its resilience to overfitting, and its efficiency in producing accurate predictions~\footnote{We leave the comparison with other classification models as future work.}.

\begin{figure}[t]
    \begin{tcolorbox}[colback=black!5!white, colframe=black!75!black, colbacktitle=black!85!black, coltitle=white]
        \small {
            This is a GitHub Issue.\\
            Repository: \textcolor{red}{$<$Repository Name$>$}\\
            Owner: \textcolor{red}{$<$Repository Owner$>$}\\
            Title: \textcolor{red}{$<$Issue Title$>$}\\
        
            --- Start of the Body ---\\
            \textcolor{red}{$<$Text of the issue$>$}\\
            --- End of the Body ---
        }
    \end{tcolorbox}
    \caption{Template of the user prompt used in the pipelines.}
    \label{fig:prompt}
\end{figure}

\begin{figure}[h]
    \centering
    \includegraphics[width=.8\linewidth]{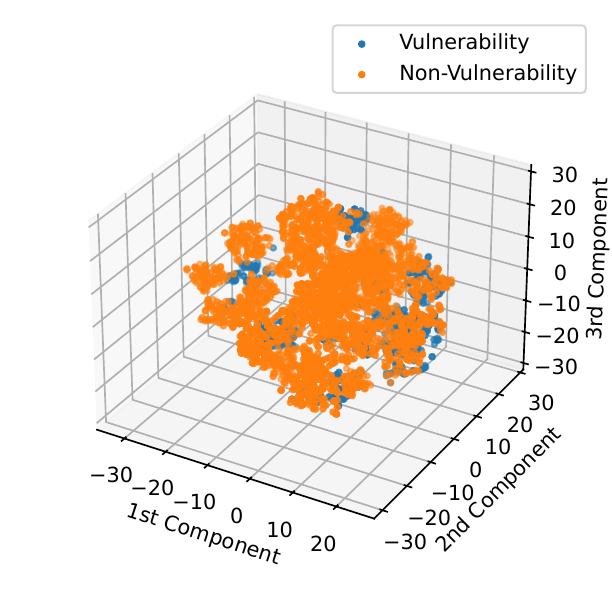}
  \caption{t-SNE plot of the dataset showing the distribution of issues. Relevant issues are marked in blue, while non-relevant ones are in orange.}
  \label{fig:tsne}
\end{figure}

\smartparagraph{Intuition: vulnerabilities are captured in the embedding space.} We run a simple analysis to explain why we believe embedding models are themselves a potentially powerful tool to predict vulnerabilities. 
Fig.~\ref{fig:tsne} shows a t-SNE projection of the distribution of our dataset of GitHub issues, revealing how issues cluster within the embedding space and indicating the potential for effective classification.

%

\subsection{LLM-based Vulnerability Detection}\label{ss:llm}

In this approach, we use LLMs to assess whether a GitHub issue is relevant for identifying a vulnerability. Compared to the binary classifier, an LLM provides several advantages, including the ability to conduct a more nuanced analysis of the given context and the capability to request a description of the problem along with a confidence score.
We prompt the LLM to engage in self-reflection and reasoning about the problem before giving an answer (\ie, a boolean), allowing it to leverage complex reasoning skills through intermediate thought processes.

Thus, for each entry of the dataset, we use a system prompt that outlines the task (reported in App.~\ref{sec:appendix:sys-prompt}) and the input shown in Fig.~\ref{fig:prompt}, asking the LLM to provide: \first a detailed explanation of the issue and any potential vulnerability identified; and \second a boolean value indicating whether the issue represents a vulnerability, based on the explanation.
We employ zero-shot learning to assess the model's ability to leverage its inherent knowledge and infer the relevance of an issue based solely on the provided textual description.

%
%

\smartparagraph{Model selection.} We selected GPT-3.5-Turbo as our target model. Although it is somewhat outdated, we specifically selected this version because its training data cutoff is September 2021, whereas newer versions (\ie GPT-4 and beyond) have a cutoff around mid-2023. This choice allows us to use the majority of our dataset for evaluation without the risk that the GitHub issues or vulnerabilities are included in the training data, which could potentially bias the results. To ensure a fairer evaluation w.r.t. the XGBoost classifier, we also fine-tuned GPT-3.5-Turbo for the specific task. For this process, we used a portion of the dataset (that is excluded from the evaluation phase, more details in Sec.~\ref{sec:evaluation}), and fine-tuned the model over three epochs.





\subsection{The Combined Approach}\label{ss:combined}

In this approach, we explore whether combining the LLM with the embedding-based classifier can improve vulnerability detection. Specifically, we aim to leverage the best of both strategies: \first the ability of LLMs to extract key information and generate detailed textual vulnerability descriptions, and \second the superior performance of a task-specific classifier. LLMs excel at summarizing textual content that may include complex technical details or specialized jargon from particular software communities or programming languages. Conversely, while the task-specific classifier could surpass the LLM in classification accuracy, it only produces a binary output and does not provide a human-readable explanation of \textit{why} a specific GitHub issue may indicate a potential vulnerability, which is much more valuable for developers and users.

\begin{figure}[t]
  \begin{center}
    \includegraphics[width=\linewidth]{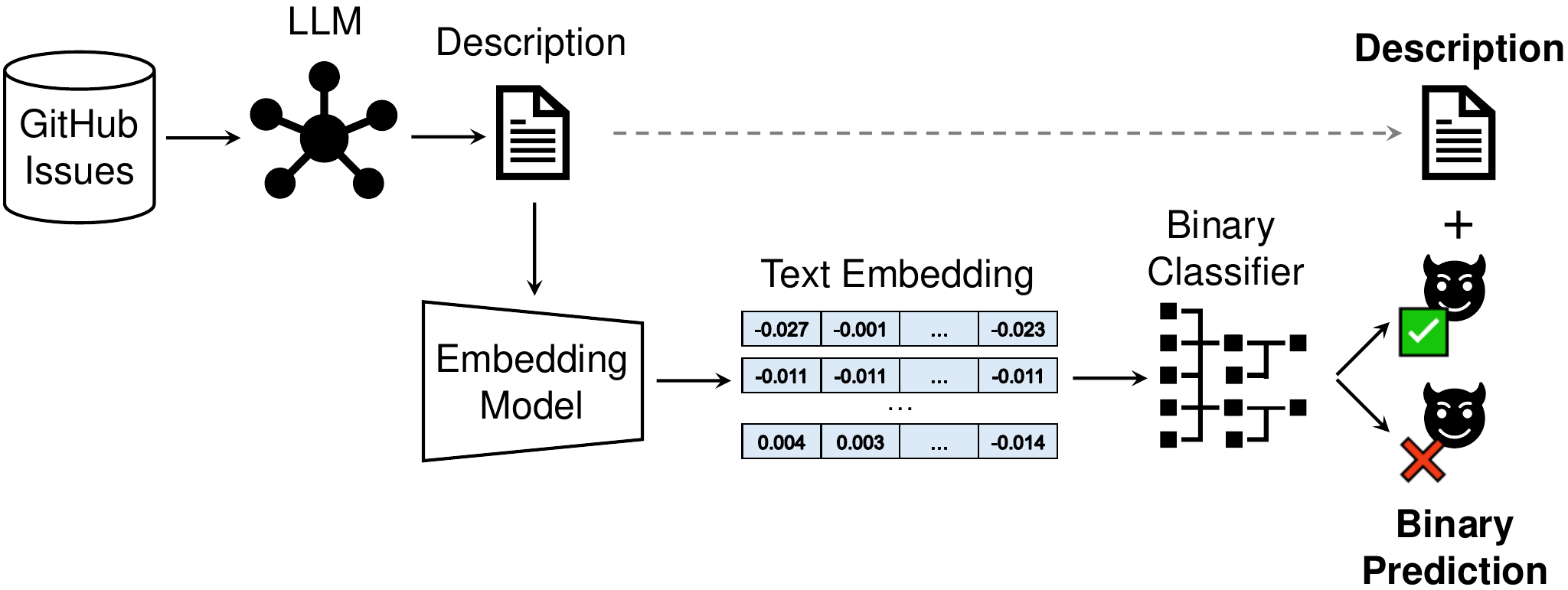}
  \end{center}
  \caption{The combined pipeline using LLM and XGBoost.}
  \label{fig:combined_solution}
\end{figure}

As depicted in Fig.~\ref{fig:combined_solution}, we start by prompting the LLM to describe and explain the input GitHub issue, summarizing the key details and highlighting the main problem discussed in the issue. This output is then converted into an embedding, which is fed into the classifier as input. The outcome of this pipeline includes both a boolean value indicating whether the GitHub issue is a potential vulnerability (from the XGBoost classifier) and a detailed description (generated by the LLM).

\smartparagraph{Model selection.} We once again use GPT-3.5-Turbo as our LLM to interpret and explain the GitHub issues, for the reasons outlined in Sec.~\ref{ss:llm}. The output is then embedded using \texttt{NV-Embed-v2}, followed by an XGBoost classifier on the embedded descriptions.

\section{Evaluation}\label{sec:evaluation}

In this section, we present a detailed analysis of the results from applying the different approaches outlined in Sec.~\ref{sec:pipelines} to the dataset we created (see Sec.~\ref{sec:dataset}).

We begin by examining the key performance metrics of the various approaches, \ie precision, recall, and F1 score, and then discuss the practical implications of each strategy. Our focus extends beyond simply detecting vulnerabilities to also providing users with insightful descriptions and potential vulnerability alerts. Thus, we then quantitatively assess the quality of the descriptions generated by the LLMs.

\smartparagraph{Comparison baseline.} We compare our proposed approaches with a baseline developed using techniques outlined in the paper ``The Tweet Advantage''~\cite{sauerwein_tweet_2018}. This baseline relies on statistical methods aimed at identifying keywords that signal vulnerability descriptions. The original method employed a regex to detect the CVE IDs within the text. For our implementation, we extended this by incorporating the following additional keywords to improve the identification of relevant issues: ``vulnerability'', ``NVD'', and ``security''.

\smartparagraph{Dataset split.} As discussed in Sec.~\ref{sec:pipelines}, we chose GPT-3.5-Turbo as our primary LLM, which has a training cutoff date of September 2021. However, our dataset includes vulnerabilities published before this date (\ie from January 2019). To ensure that the evaluation data does not overlap with the model's training data, we excluded all issues and vulnerabilities created before the cutoff date, which accounted for 15\% of the total dataset. An additional 3\% of issues were excluded to prevent data contamination, as these issues were created before the cutoff date, but the vulnerabilities were detected afterward.
We then divided the remaining data into two splits: \textit{post-cutoff training} and \textit{post-cutoff test}. The post-cutoff training split, comprising 49\% of the total dataset, was used exclusively for training the XGBoost classifiers and fine-tuning the LLM. The post-cutoff test, making up 33\% of the total dataset, is solely employed for model evaluation. 


Note that the dataset composition is imbalanced, with a larger proportion of issues not linked to any vulnerabilities (\ie 3,611 non-vulnerability entries vs. 443 vulnerability entries). We intentionally did so to reflect a realistic scenario where vulnerability-related issues are less frequent.


\begin{table}[b]
\caption{Comparison among the different approaches.}
\label{tab:comparison}
\centering
\resizebox{\linewidth}{!}{%
\begin{tabular}{l c c c c}
\multicolumn{1}{c}{\textbf{Model}} & \multicolumn{1}{c}{\textbf{Class}} & \multicolumn{1}{c}{\textbf{Precision}} & \multicolumn{1}{c}{\textbf{Recall}} & \multicolumn{1}{c}{\textbf{F1 score}} \\
\toprule
Baseline~\cite{sauerwein_tweet_2018} & No Vuln. & 0.90 & 0.96 & 0.93 \\
{} & Vuln.  & 0.45 & 0.24 & 0.31 \\
\midrule
LLM-only (GPT-3.5) & No Vuln. & 0.97 & 0.76 & 0.85 \\
{} & Vuln. & 0.33 & 0.85 & 0.48 \\
\midrule
LLM-only (GPT-3.5 Fine-Tuned) & No Vuln. & 0.96 & 0.83 & 0.89 \\
{} & Vuln. & 0.37  & 0.73 & 0.49 \\
\midrule
OpenAI Embeddings + XGBoost & No Vuln. & 0.93 & 0.95 & 0.94 \\
{} & Vuln. & 0.58 & 0.53 & 0.55 \\
\midrule
Nvidia Embeddings + XGBoost & No Vuln. & 0.93 & 0.97 & \textbf{0.95} \\
{} & Vuln. & 0.70 & 0.50 & 0.58 \\
\midrule
Combined & No Vuln. & 0.94 & 0.96 & \textbf{0.95} \\
(GPT-3.5 + Nvidia Emb. + XGBoost) & Vuln. & 0.66 & 0.55 & \textbf{0.60} \\
\end{tabular}
}
\end{table}


\subsection{Performance Evaluation}

We begin by analyzing the key performance metrics of the various approaches. Table~\ref{tab:comparison} reports precision, recall, and F1 score for the two main classes of interest, vulnerabilities (\ie Vuln.) and non-vulnerabilities (\ie No Vuln.), as evaluated on the post-cutoff test split of our dataset.

\smartparagraph{Transformer-based approaches outperform statistical methods.} We start by briefly examining the performance of all our Transformer-based approaches compared to the regex-based methodology of the baseline. Our proposed approaches retain a higher F1 score, with the combined pipeline (last row of the table) performing \textit{$\sim$2$\times$ better} in detecting vulnerabilities  compared to the baseline. The results confirm our hypothesis that Transformer-based models are more capable of identifying vulnerabilities, thanks to their deeper semantical understanding of the textual data, underscoring their potential for more accurate and reliable early vulnerability detection in real-world scenarios.

\smartparagraph{The LLM-based approach is not the best performing, but newer models could improve results with larger datasets.} Despite LLMs have demonstrated exceptional performance across various domains, they do not perform best in the task of vulnerability detection (see second and third row of Table~\ref{tab:comparison}). We speculate this is primarily due to two factors: \first the chosen model version (GPT-3.5-Turbo) is somewhat outdated and lacks the advanced reasoning capabilities present in newer versions. However, using more recent models was not feasible due to their training cutoff date, as it would reduce the dataset to fewer than 100 items, which is insufficient for effectively validating our approach; and \second by handpicking some of the issues in the dataset, we observed that some refer to problems such as buffer overflows, heap overflows, or program crashes. In these cases, the LLM tends to flag such issues as potential vulnerabilities due to their severity. However, since these issues are not linked to any CVE entries, they are marked as incorrect responses during the evaluation process. To make a fairer evaluation, we \textit{fine-tuned} GPT-3.5-Turbo using the \textit{same training set} as the XGBoost classifier. As expected, the F1 score slightly improved, but the previously mentioned issues continued to persist. Indeed, this is not a definitive conclusion, as we believe that newer models could achieve better performance if evaluated with a sufficiently large post-cutoff dataset, which remains our future work. Moreover, also asking the LLM to be more or less conservative with the answers, may affect the results, which we leave for future studies. 

\smartparagraph{Embedding models can detect vulnerabilities with high accuracy.} Since the embedding models are also based on the Transformer architecture, they can focus on various parts of the input sequence, effectively capturing contextual relationships and semantic details. Combining these embeddings with XGBoost resulted in high accuracy for detecting vulnerabilities, as shown in Table~\ref{tab:comparison} (fourth and fifth rows). Surprisingly, the open-source \texttt{NV-Embed-v2} produced higher-quality embeddings than \texttt{text-embedding-3-large} from OpenAI. 
As mentioned in Sec.~\ref{sec:pipelines}, the \texttt{NV-Embed-v2} model is based on \texttt{Mistral-7B-v0.1}, which has a training cutoff around August 2021, aligning with the timeframe of the evaluation dataset and ensuring the results are valid.
This finding has several important implications: \first it demonstrates that open-source models can outperform proprietary ones, which is beneficial for scenarios where it is not possible to access external APIs; \second open-source models are typically smaller, allowing deployment and execution on commodity hardware, thereby reducing operational costs; and \third the overall pipeline remains highly efficient, as both the embedding model and XGBoost have significantly lower computational costs compared to LLMs. This efficiency makes it an ideal and practical solution for this task, since it requires analyzing large volumes of textual data from diverse sources in real time. 
However, the main limitation of using binary classifiers is their inability to provide a textual description of the issue, which is instead valuable for developers and users.





\smartparagraph{The combined model achieves the best performance while providing insightful vulnerability descriptions.}
%
As already mentioned, the combined pipeline achieves an F1 score for vulnerability detection that is twice as high as the baseline (see the last row of Table~\ref{tab:comparison}). It effectively combines superior classification accuracy (using \texttt{NV-Embed-v2} and XGBoost) with the capability to produce informative descriptions of potential vulnerabilities. Although the improvement in classification performance over the embedding-only model is moderate, this integration allows for more detailed analysis of the issues through embeddings, boosting overall performance. We expect our system to scan millions of GitHub issues in real time, therefore efficiency and costs are important factors to consider. The combined approach stands out as the most efficient and cost-effective option, leveraging an economical LLM (GPT-3.5-Turbo) alongside a compact embedding model and XGBoost classifier that can be easily deployed on commodity hardware. Additionally, GPT-3.5-Turbo can be replaced with a similar open-weights LLM (\eg Mistral-7B), further reducing operational costs.
Indeed, we hypothesize that further improvements could be realized by using an LLM with advanced reasoning capabilities to generate the descriptions. In our evaluation, we used previously discovered vulnerabilities, so it was essential to consider the training cutoff to prevent biases. However, in a real-time deployment, it would be impossible for the LLM to include newly created GitHub issues or CVEs, allowing the use of more powerful models. We plan to explore this additional evaluation in future work.


\begin{figure}[b]
    \centering
    \includegraphics[width=\linewidth]{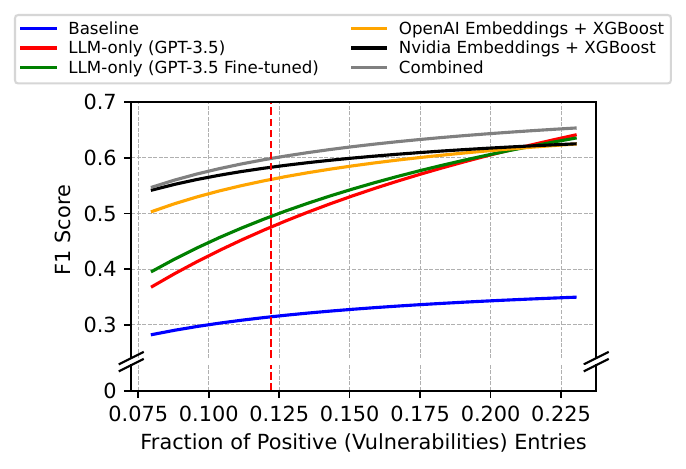}
    \caption{Sensitivity analysis. The vertical dashed red line represents the ratio used to train the XGBoost classifier.}
    \label{fig:sensitivity}
\end{figure}

\smartparagraph{Sensitivity analysis}. 
We trained our XGBoost classifier on a dataset that consisted of 12.2\% of positive elements (\ie vulnerabilities), and tested it on a dataset with the same ratio. Now, we aim to assess how the F1 score varies with changes in this ratio without retraining the XGboost classifier, \ie we do not modify the classifier but only adjust the proportion of positive elements over the total dataset.
To calculate the F1 score for different ratios, we leverage the recall metric on both positive and negative samples since the recall metric does not depend on the ratio of positive to negative samples. 

Fig.~\ref{fig:sensitivity} shows the F1 score for vulnerability detection (y-axis) as the proportion of positive instances in the test dataset varies between 8\% and 20\%. We skew the analysis towards higher values as we hypothesize that most vulnerabilities after the cutoff date may still require to be reported through CVEs. This will inevitably increase the ratio once these vulnerabilities are reported. Our results demonstrate that the combined approach remains robust and outperforms other methods. Additionally, we notice that the performance of LLM-based approaches improves as the ratio increases, which aligns with the observation that LLMs tend to be ``conservative'' and flag anything potentially suspicious as a vulnerability.

We leave the dynamic retraining of the classifier based on the observed precision and recall metrics as an open question. This task is non-trivial as we observe a sharp difference, for instance, in the ratio of positive to negative samples in the pre- and post-cutoff datasets. 

\subsection{Quality of the LLM-Generated Descriptions}

\begin{figure}
    \centering
    \subfigure[GPT-3.5-Turbo.]{
        \includegraphics[width=.8\linewidth]{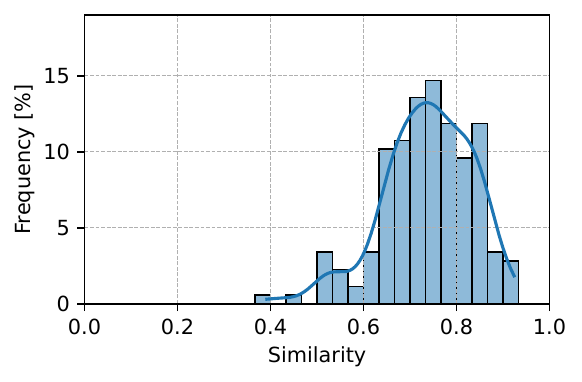}
        \label{fig:similarity:gpt3.5_similarity}
    }    
    \subfigure[GPT-3.5-Turbo (Fine-tuned).]{
        \includegraphics[width=.8\linewidth]{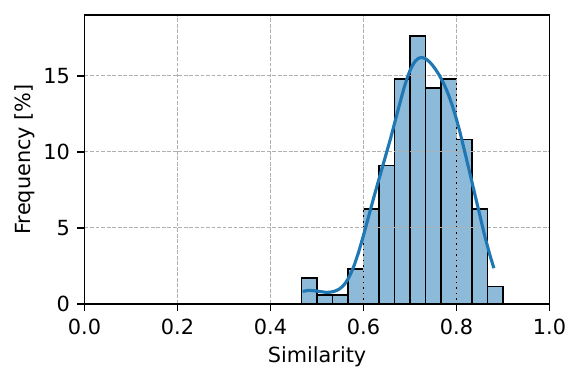}
        \label{fig:similarity:gpt3.5_ft_similarity}
    }
    \caption{Similarity distribution between the vulnerability description in the CVE record and the one generated by the LLMs, measured using cosine similarity.}
    \label{fig:similarity}
\end{figure}

One significant advantage of utilizing LLMs is their capability to generate insightful descriptions and identify potential vulnerability alerts. In the following analysis, we assess the quality of these generated descriptions to determine whether they provide sufficient information to indicate a vulnerability. We compare the descriptions produced by GPT-3.5-Turbo and its fine-tuned version with the official descriptions in the CVE records.
To evaluate the similarity between the two descriptions, we \first generate their embeddings using again the \texttt{NV-Embed-v2} model, which converts the texts into dense vector representations to capture semantic meaning, and \second compute the cosine similarity between the vectors. The results are presented in Fig.~\ref{fig:similarity}, which displays a histogram of the distribution of similarity scores. Notably, the distribution of descriptions generated by the fine-tuned GPT-3.5 (Fig.~\ref{fig:similarity:gpt3.5_ft_similarity}) shows a higher density at higher similarity levels compared to the base GPT-3.5 (Fig.~\ref{fig:similarity:gpt3.5_similarity}).

Our results provide two important insights: \first LLMs are capable of generating vulnerability descriptions that are semantically similar to the official descriptions. This confirms that LLMs excel in summarization and can offer valuable details for future, undisclosed vulnerabilities; and \second fine-tuning enhances the performance of the base LLM, allowing GPT-3.5 to generate descriptions that more closely match those found in CVE records.


\section{Conclusion}

In this paper, we reveal the challenge of vulnerability detection prior to their official disclosure date and propose using Transformer-based models as a solution. By building a dataset that links GitHub issues with their corresponding CVE references, and evaluating three different Transformer-based approaches, we demonstrate that these models can simplify and effectively address this task. This work contributes by \first introducing a novel research direction: detecting early indicators of potential CVE vulnerabilities, and \second proposing a framework for automated vulnerability detection that balances accuracy, explainability, and computational efficiency. By addressing these key challenges, our work provides valuable tools to help developers identify and mitigate vulnerabilities in their code bases, significantly enhancing the security of open-source software ecosystems.

\section*{Acknowledgments}

We would like to thank the anonymous reviewers for their insightful comments and suggestions on this paper. This work has been partially supported by Vinnova (the Sweden's Innovation Agency), the Swedish Research Council (agreement No. 2021-04212), Knut and Alice Wallenberg Foundation (Wallenberg Scholar Grant for Prof. Dejan Kosti\'c), and KTH Digital Futures.

\bibliographystyle{IEEEtran}
\small
\bibliography{references}

\newpage
\appendices
\section{LLM System Prompt}\label{sec:appendix:sys-prompt}

Below, we provide the system prompt used to instruct the LLM (\ie GPT-3.5-Turbo) in performing vulnerability classification.

\begin{tcolorbox}[colback=black!5!white, 
    colframe=black!75!black, 
    colbacktitle=black!85!black,
    coltitle=white]
You are a cybersecurity assistant tasked with identifying potential vulnerabilities by analyzing GitHub issues. Your goal is to review each issue and determine whether it indicates a security vulnerability. Provide a detailed description of the issue and a confidence score of how much you are confident about the vulnerability.
\\\\
In addition to identifying security vulnerabilities, you should also recognize cases where the issue is not a vulnerability. These may include failing tests, minor bugs, or issues related to functionality that do not present security risks.
\\\\
Please format your response in JSON with the following fields. DO NOT add any additional text to the response:
\\\\
description: Describe and reason about the issue. Explain if you detect any potential vulnerability or not.
\\\\
vulnerability\_detected: A boolean indicating whether the issue is relevant (true) or not (false). This should be based on the explanation in description.
\end{tcolorbox}
\end{document}